\documentclass[preprint,floatfix
,aps,prd,superscriptaddress]{revtex4}

\usepackage[utf8]{inputenc}
\usepackage{graphicx}
\usepackage{tikz}
\usepackage{amssymb,amsmath,mathtools,comment,mathrsfs,hyperref}
\newcommand{\ket}[1]{|#1\rangle}
\newcommand{\bra}[1]{\langle #1|}

\newcommand{\rs}{\rho_{\!_S}}
\newcommand{\sgn}{\operatorname{sgn}}
\begin{document}

\title{
Thermal effect in a causal diamond: open quantum systems approach}
\author{A. Chakraborty}
\affiliation{Department of Physics, University of Houston. Houston, Texas 77024-5005, USA}
\author{H. E. Camblong}
\affiliation{Department of Physics and Astronomy, University of San Francisco. San Francisco, California 94117-1080, USA}
\author{C. Ord\'o\~nez}
\affiliation{Department of Physics, University of Houston. Houston, Texas 77024-5005, USA}

\date{\today}
\begin{abstract}
A static observer with a finite lifetime has causal access to only a limited region of spacetime known as the causal diamond. The presence of an apparent horizon in the causal diamond, due to the observer's finite lifetime, is the origin of an Unruh-like thermal effect. Thus, even though the observer is static and the background is flat, the finite-lifetime observer experiences a thermal bath in the Minkowski vacuum. In this article, we provide an open quantum systems approach that yields a complete thermal characterization via the observer's steady-state density matrix, which is shown to be thermal with a temperature inversely proportional to its lifetime. This associated diamond temperature agrees with the established result derived from other methods. Moreover, our approach is particularly useful for designing entanglement harvesting protocols in the causal diamond. In addition, we introduce an insightful procedure that defines diamond coordinates using conformal transformations, and which leads to a more direct derivation of the thermal properties. 
\end{abstract}


\maketitle

\section{Introduction}
The concept of the vacuum state is observer dependent in quantum field theory in curved spacetime. This was illustrated by Stephen Hawking in his seminal papers \cite{hawking1,hawking2,hawking3} on thermal radiation emitted by black holes. In another seminal paper \cite{unruh76}, William Unruh showed that the Hawking effect is equivalent to a uniformly accelerated observer detecting thermal particles in the Minkowski vacuum, and the temperature is proportional to the acceleration of the observer. The temperature $T_U = a/2\pi$ (in natural units $\hbar=G=c=k_B=1$) is known as the Unruh temperature and the effect is known as the Unruh effect. The Unruh effect showed explicitly that the notion of the presence or absence of particles in a quantum state is related to the motion of a detector. This was corroborated further by Davies, Fulling, and others in a series of papers \cite{wald75,parker75,ufd76,davies78,df77}. The origin of this thermality is attributed to the presence of an event horizon (in the case of a black hole) or an apparent horizon (in the case of a uniformly accelerated observer) \cite{israel76}. Due to the presence of a horizon, the observer can only access a certain region of the spacetime. Thus, any measurement in the frame of the observer should involve an average over the degrees of freedom associated with the inaccessible region. This averaging yields a mixed state for the observer; and in the specific case of an accelerated observer, it yields a thermal state. The contribution of the horizon to determination of the thermality was further illustrated by the conformal field theory approach, where the central charge of the Virasoro algebra was shown to be related to the Hawking temperature \cite{carlip1,carlip2,birminghamsen,cardy}. Another near-horizon approach used the conformal symmetry of the fields near the horizon to show that it is possible to derive the area-entropy formula for generic black holes by statistical mode counting \cite{nhcamblong,nhcamblong-sc,nhcamblong-tightness}. In a recent series of papers, using a quantum optics approach, it was shown that the presence of the horizon is responsible for creating the thermal atmosphere in static and stationary black holes \cite{chakraborty1,chakraborty2,chakraborty3,chakraborty4}. \\

\begin{figure}[h]
    \centering
    \includegraphics[width=0.3\linewidth]{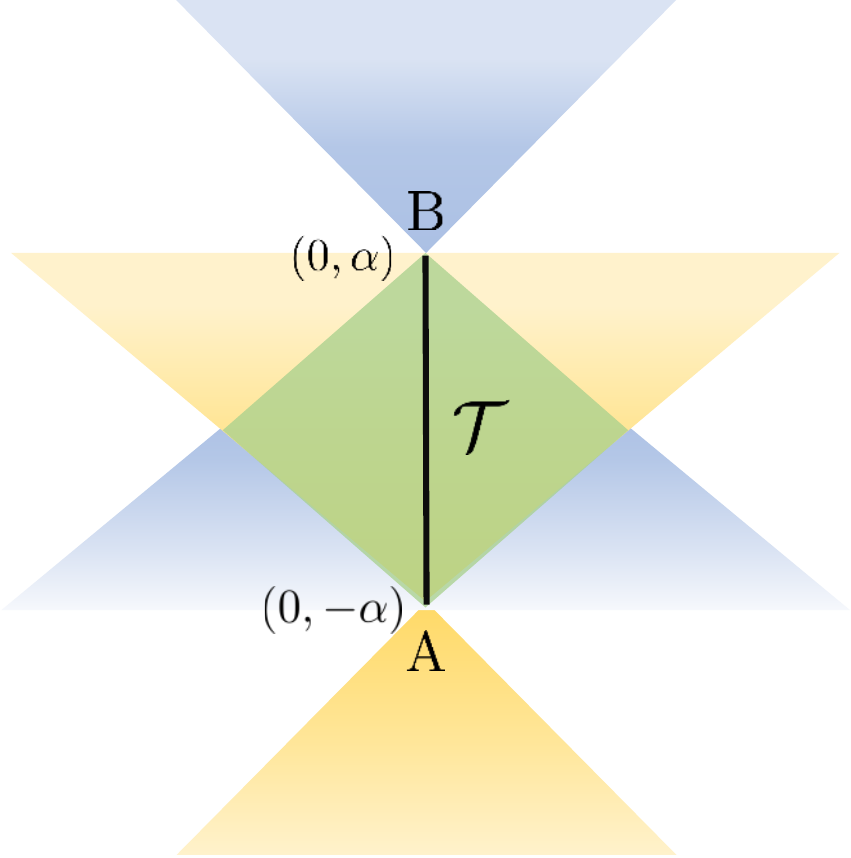}
    \caption{Here, A is the point of birth of the observer and B is the point of death. So the length of AB is the lifetime of the observer $\mathcal{T}=2\alpha$. The intersection of the future light cone of A and the past light cone of B is the causal diamond (shaded region in green). The boundary of this diamond region acts as a horizon for a diamond observer.}
    \label{fig:causal-diamond-schematics}
\end{figure}

In all the approaches mentioned above, the Unruh effect and the Hawking effect are always associated with an accelerated system---either the observer or the field. However, in 2003, Martinetti and Rovelli, in their seminal paper \cite{martinetti-1}, showed that a finite-lifetime observer can detect thermal particles in the Minkwoski vacuum despite being static. In essence, the birth and death of the finite-lifetime observer restrict the observer's causal access to a finite region of the whole Minkowski spacetime. This is the region causally connected with the observer with a given lifetime; thus, it is the intersection between the future light cone of the birth event and the past light cone of the death event. Because of its diamond-shaped appearance, this region is called the causal diamond (Fig.~\ref{fig:causal-diamond-schematics}). Due to the presence of the apparent horizons in the causal diamond structure, the finite-lifetime observer will not perceive the Minkowski vacuum as a vacuum state. Martinetti and Rovelli showed that the static observer experiences a variable temperature in the Minkowski vacuum. The physical meaning of this result was clarified in a series of papers \cite{su-ralph-1,su-ralph-2,light-cone,jacobson} in the context of black holes, AdS spacetime, and Unruh-deWitt detectors. We particularly focus on the detector approach \cite{su-ralph-1,su-ralph-2}, in which the finite-lifetime observer can be treated as a two-level system with the energy gap between the two-levels fixed in the system's reference frame. These two-level systems, known as the diamond observers, perceive a different vacuum state than the Minkowski vacuum and the excitation rate of these observers in the Minkowski vacuum is thermal in nature. This temperature $T_D = 2/\pi \mathcal{T}$ is known as the diamond temperature and is inversely proportional to the lifetime $\mathcal{T}$ of the observer. One can also show that the annihilation and creation operators of a diamond observer and a Minkowski observer are related by a nontrivial Bogolyubov transformation \cite{su-ralph-1}, which is a trademark of the inequivalence of the diamond and Minkowski vacuum. Moreover, the expression for the diamond temperature satisfies the requirement that the static observer becomes a Minkowski observer as the lifetime increases to infinity.

Given the physical relevance of this result, it is worthwhile to explore other derivations of the thermality of the state of the finite-lifetime observer. In this paper, we focus on a derivation using the open quantum systems approach, where the diamond observers are treated as the system and the scalar field configuration is treated as the environment. This derivation is important for the following three purposes: (i) it illustrates the role of decoherence in determining the diamond temperature, (ii) it paves the path for designing entanglement harvesting protocols between two diamond observers, and (iii) it can be helpful in simulating the evolution of the system in a quantum computer. The role of decoherence in determining the thermality of a system has been previously illustrated in the literature for accelerated observers and black holes \cite{benatti,scully,yu-zhang-1}. In our case, we show that the system-environment interaction can lead to thermality for a static observer in a causal diamond. This approach also shows the importance of the role played by tracing out the environment degrees of freedom to generate a reduced thermal density matrix for the diamond observer. On the other hand, the interaction of two observers with the same environment can lead to enhancement or destruction of entanglement between the two systems. The open quantum systems approach offers a more direct way to evaluate the evolution of the entanglement between the two particles, as has been illustrated in a number of papers \cite{benatti,huyu-1,huyu-2,huyu-3,yu-zhang-2,yu-zhang-3}. Open quantum systems have also been simulated successfully in quantum computers, and some efficient quantum algorithms have been used to solve specific cases of the Lindblad equation \cite{oqs-qc-1,oqs-qc-2,oqs-qc-3,oqs-qc-4}. Therefore, it is worth studying the open quantum system approach for causal diamonds as it relates to the feasibility of simulating it in a quantum computer in the near future.\\

For an accelerated system, the open quantum system approach has been used for a broad class of Hamiltonians and interactions \cite{benatti,yu-zhang-1}. For the purposes of our paper, we consider a similar approach for causal diamonds and illustrate the mechanism of decoherence for the case where the Hamiltonian is directed along the $z$-direction and the interaction between the diamond observer and the field is a monopole interaction with a single scalar field. In addition, we define a coordinate system suitable for the diamond observer. There are a few variants of the diamond coordinates that have appeared in the literature over the years \cite{martinetti-2,jacobson,su-ralph-1}. In this article, we provide a general method of defining the diamond coordinates using conformal transformations, which is more direct than the one used in the original derivation by Martinetti and Rovelli \cite{martinetti-1,martinetti-2}, and can be traced back to similar forms in the literature~\cite{jacobson,casini}.

This paper is organized as follows. In Sec.~\ref{sec:diamond-coordinates}, we discuss the generic procedure for defining the diamond coordinates, as well as particular cases, including the choice used by Martinetti and Rovelli; and we consider the trajectory of a static observer in the diamond coordinates. In Sec.~\ref{sec:udw-detector-scaled}, we define our model of an Unruh-DeWitt detector. In Sec.~\ref{sec:lindblad}, we briefly outline the open quantum systems approach, apply the Lindblad equation to our system, and find the evolution of the density matrix of the system. In Sec.~\ref{sec:steady-state}, we determine the steady-state density matrix of the system and discuss its thermal nature. We conclude the paper with a discussion of our findings and directions for future work in Sec.~\ref{sec:conclusions}.
The appendices show how to evaluate the integral critical to thermality 
and the details of a generalized comparative framework for the Bloch-vector form of the Lindblad equation.

\section{Diamond Coordinates}\label{sec:diamond-coordinates}
\begin{figure}[t]
    \centering
    \includegraphics[width=0.8 \linewidth]{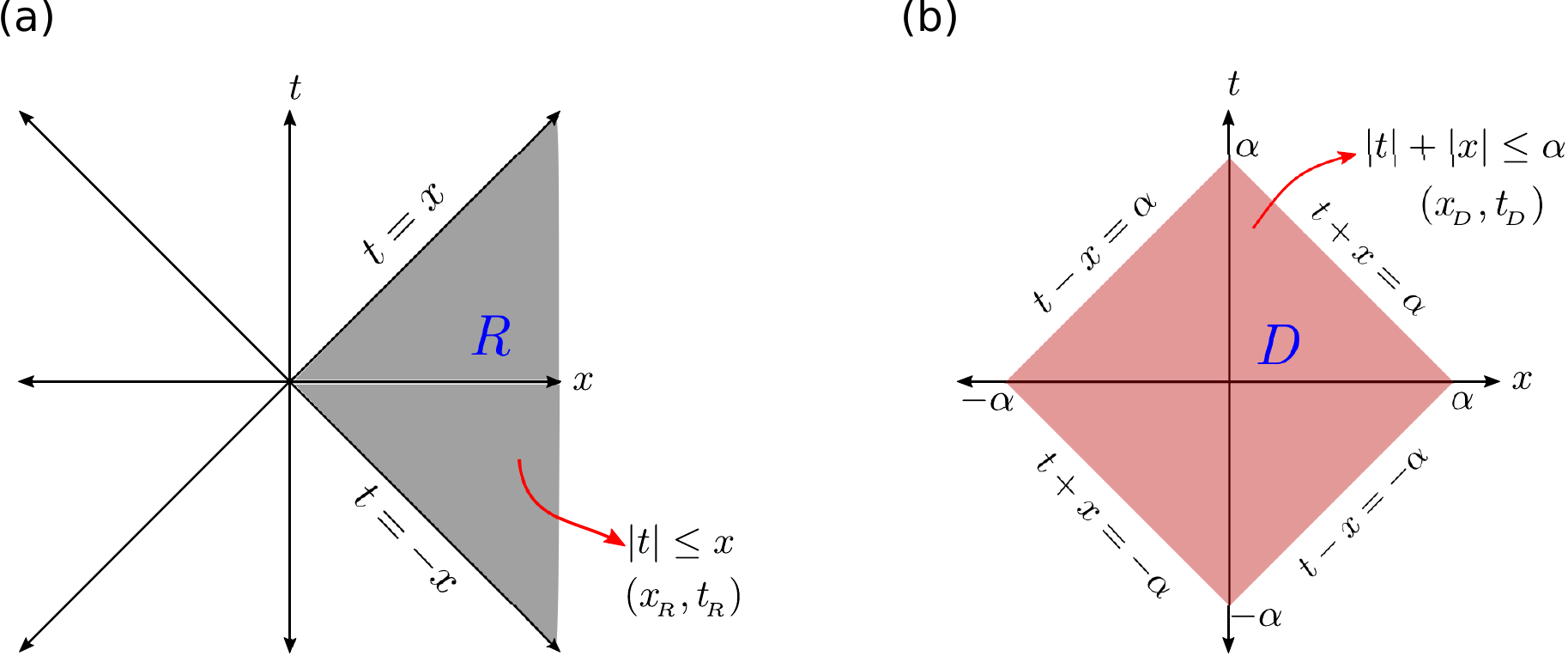}
    \caption{(a) The shaded region of the (1+1)-dimensional Minkowski spacetime shows the right Rindler wedge, which we call $R$. (b) The diamond-shaped shaded region $D$ centered at the origin is the causal diamond for a static observer at the origin with finite lifetime from $-\alpha$ to $\alpha$.}
    \label{fig:R&D region}
\end{figure}

In the introduction, we defined the causal diamond to be the region bounded by the intersection of the past light cone of the death event and the future light cone of the birth event of a finite-lifetime observer (Fig.~\ref{fig:causal-diamond-schematics}). In this section, we are going to introduce a generic procedure to define diamond coordinates, namely, coordinate systems suitable for the associated diamond observer with a finite lifetime. Let $(\xi,\eta)$ denote diamond coordinates that cover the whole range of the causal diamond region. Such coordinates are required for the quantization of the field and to describe any causally accessible event from the observer's perspective. However, this coordinate chart is not unique, as will be obvious from the procedure outlined below; in effect, in the literature, there exist a few alternative versions of diamond coordinates. Thus, we propose an underlying principle for defining such a coordinate chart: the existence of a nonunique conformal mapping between the right Rindler wedge $R\coloneqq\{(x,t): |t|\leq x \text{ and }  x\geq 0\}$ and the diamond region $D\coloneqq \{(x,t): |t|+|x|\leq\alpha\}$, where $2\alpha$ is the lifetime of the observer. In what follows, we will refer to $2\alpha$ as the size of the diamond; and will denote the $(x,t)$ points in Minkowski space restricted to region $R$ as $(x_{\!_R},t_{\!_R})$, and the $(x,t)$ points restricted to the diamond region as $(x_{\!_D},t_{\!_D})$. A visual representation of these regions can be seen in Fig.~\ref{fig:R&D region}.

Once such a mapping between $(x_{\!_R},t_{\!_R}) \leftrightarrow (x_{\!_D},t_{\!_D})$ is established, we can 
use the relation between the Rindler wedge $R$ and Rindler coordinates to define a coordinate chart $(\xi,\tau)$ suitable for the diamond observer,
\begin{equation}
    \tau = \alpha\tanh^{-1}\left(\frac{t_{\!_R}}{x_{\!_R}}\right) \qquad\qquad 
    \xi = \frac{\alpha}{2}\ln\left[\alpha^{-2}(x_{\!_R}^2 - t_{\!_R}^2)\right]\;, \label{eq:diamond-rindler-pre}
\end{equation}
where $\xi,\tau \in (-\infty,\infty)$, i.e. they cover the whole $\mathbb{R}^2$ plane. Equation~(\ref{eq:diamond-rindler-pre}) gives the inverse transformation of the relation between the Rindler coordinates and Minkowski coordinates,
\begin{equation}
    t_{\!_R} = \alpha\, e^{\xi/\alpha}\sinh(\tau/\alpha)\;,\qquad \qquad x_{\!_R} = \alpha\, e^{\xi/\alpha}\cosh(\tau/\alpha)\;. \label{eq:rindler-minkowski}
\end{equation}
Here, $\xi=$ constant denotes a uniformly accelerated observer with acceleration $(\alpha\, e^{\xi/\alpha})^{-1}$. The sequence of steps above establishes a mapping from the diamond region to the $\mathbb{R}^2$ plane via the path $(x_{\!_D},t_{\!_D}) \rightarrow (x_{\!_R},t_{\!_R}) \rightarrow (\xi,\tau)$. The $(\xi,\tau)$ coordinate system is suitable for the diamond observer because each possible value of $(\xi,\tau)$ is connected to a unique spacetime point in the diamond via this mapping. For future convenience in Sec.~\ref{sec:udw-detector-scaled}, we rescale the time coordinate $\tau$ in $(\xi,\tau)$ by a factor of 1/2, i.e., we define $\eta =  \tau/2$. This rescaling is one-to-one and the new coordinate chart $(\xi,\eta)$ also covers the $\mathbb{R}^2$ plane. In the new coordinate system we have
\begin{equation}
    \eta = \frac{\alpha}{2}\tanh^{-1}\left(\frac{t_{\!_R}}{x_{\!_R}}\right) \qquad\qquad 
    \xi = \frac{\alpha}{2}\ln\left[\alpha^{-2}(x_{\!_R}^2 - t_{\!_R}^2)\right]\;, \label{eq:diamond-rindler}
\end{equation}

The critical step in the definition of the diamond coordinates is the conformal mapping between the regions $R$ and $D$. The nonuniqueness of this mapping is the reason for the appearance of different diamond coordinate charts for $D$ in the literature. In the following, we first provide a generic procedure to generate such diamond coordinate charts, and we consider the simplest possible choice that we will later use for the open quantum systems approach. In addition, we will show that this general procedure also yields the coordinate chart used by Martinetti and Rovelli \cite{martinetti-1,martinetti-2}.
 
\subsection{General procedure for the conformal mapping} \label{subsec:general-conformal}
The building blocks for defining the mapping between $D$ and $R$ are the three following conformal transformations:
\begin{enumerate}
    \item Special conformal transformation $K(\rho)$ defined as 
    \begin{equation}
    K(\rho)\; x^\mu = \frac{x^\mu - b^\mu (x\cdot x)}{1-2 (b\cdot x) + (b\cdot b) (x\cdot x)} \label{eq:sct_MR}
    \end{equation}
    where $b^\mu = (0,-\rho,0,0)$ is directed along the $x$-axis in four-dimensional spacetime. The dot product represents the inner product in the Minkowski space with metric signature $(-,+,+,+)$. The special conformal transformation $K(\rho)$ maps the right Rindler wedge $R$ to a diamond of size $1/\rho$ centered at $(1/2\rho,0)$ in the $(x,t)$ plane. In general, this type of special conformal transformation leaves the structure of the light cone invariant and maps a diamond region into another diamond region.
    \item Scaling transformation $\Lambda(l)$ defined as:
    \begin{equation}
        \Lambda(l) \;x^\mu = lx^\mu\;.
    \end{equation}
    This transformation is useful to rescale the size of the diamond. Applying this transformation on a diamond of size $1/\rho$ would yield a diamond of size $l/\rho$.
    \item Spatial translation $T(a)$ defined as
    \begin{equation}
        T(a) \;x = x+a \;.
    \end{equation}
    This transformation can be used to bring the center of the diamond to the origin.
\end{enumerate}
The general procedure is a composition of these building blocks. The first step is to map the right wedge $R$, which can be thought as an infinite-sized diamond, to a finite-sized diamond using the special conformal transformation $K(\rho)$. Once the diamond is brought to a finite size, one can use rescaling and translation as intermediate steps to bring the diamond to the desired size and center it at the origin. One can choose any finite value of $\rho$ for the special conformal transformation and use multiple rescaling and translation to get the desired diamond size. This gives rise to different coordinate charts. Here, one should note that the generators of the three transformations $K(\rho)$, $\Lambda(l)$, and $T(a)$ do not commute. In fact, the generators satisfy the $sl(2,\mathbb{R})$ algebra. So, changing the order of the transformations will lead to a different mapping.

The simplest way~\cite{jacobson} to map the region $R$ to a diamond $D$ of size $2\alpha$ is to use the composite map
\begin{equation}
    T(-\alpha) \circ K\left(\frac{1}{2\alpha}\right) \;. \label{eq:jacobson-diamond}
\end{equation}
Here, one maps the region $R$ to a diamond of the desired size $2\alpha$ and then use translation to bring the center of the diamond to the origin. The scaling transformation is not used in this case as we directly map $R$ to a diamond of the desired size. One can use the expression for $K(\rho)$ and $T(a)$ to obtain the mapping between $(x_{\!_R},t_{\!_R}) \rightarrow (x_{\!_D},t_{\!_D})$ explicitly: 
\begin{align}
    t_{\!_D} &= \frac{t_{\!_R}}{1+(x_{\!_R}/\alpha) + (x_{\!_R}^2-t_{\!_R}^2)/4\alpha^2} \;, \\
    x_{\!_D} &= \frac{(x_{\!_R}^2-t_{\!_R}^2)/4\alpha - \alpha}{1+(x_{\!_R}/\alpha)+(x_{\!_R}^2-t_{\!_R}^2)/4\alpha^2} \;.
\end{align}
However, one needs the inverse transformation $(x_{\!_D},t_{\!_D}) \rightarrow (x_{\!_R},t_{\!_R})$ to define the diamond coordinates in Eq.~(\ref{eq:diamond-rindler}). The inverse transformation can be obtained easily using the inverted composite map $K^{-1}(1/2\alpha)\circ T(\alpha)$, where $K^{-1}(1/2\alpha)= K (-1/2\alpha) $, which yields the following coordinate chart,
\begin{align}
    t_{\!_R} &= \frac{4\alpha^2 t_{\!_D}}{(x_{\!_D}+\alpha)^2 - t_{\!_D}^2 - 4\alpha x_{\!_D}}
     \label{eq:rindler-diamond-jacobson_t} \\
    x_{\!_R} &= \frac{4\alpha^2(x_{\!_D}+\alpha) - 2\alpha[(x_{\!_D}+\alpha)^2-t_{\!_D}^2]}{(x_{\!_D}+\alpha)^2 - t_{\!_D}^2 - 4\alpha x_{\!_D}} \label{eq:rindler-diamond-jacobson_x}
    \; .
\end{align}
In the remainder of the paper, we are going to use this particular mapping to define the diamond coordinates due to its geometric simplicity. However, it is obvious that this is only one particular way to define the diamond coordinates. For completeness, we will also rederive the original diamond coordinates by Martinetti and Rovelli below.
\begin{figure}[t]
    \centering
    \includegraphics[width=1.0\linewidth]{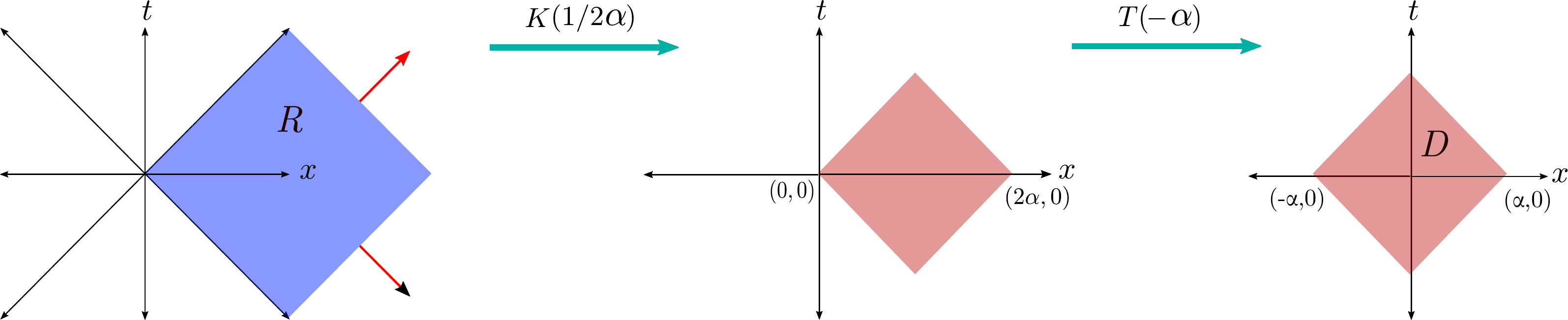}
    \caption{This figure shows the composite map, Eq.~(\ref{eq:jacobson-diamond}). The right Rindler wedge $R$ can be thought of as a diamond of infinite size, i.e., a diamond whose edges are taken to infinity. }
    \label{fig:Rindler-to-diamond-jacobson}
\end{figure}

\subsection{Diamond coordinates by Martinetti and Rovelli} \label{subsec:martinetti-rovelli}
\begin{figure}[t]
    \centering
    \includegraphics[width=0.7\linewidth]{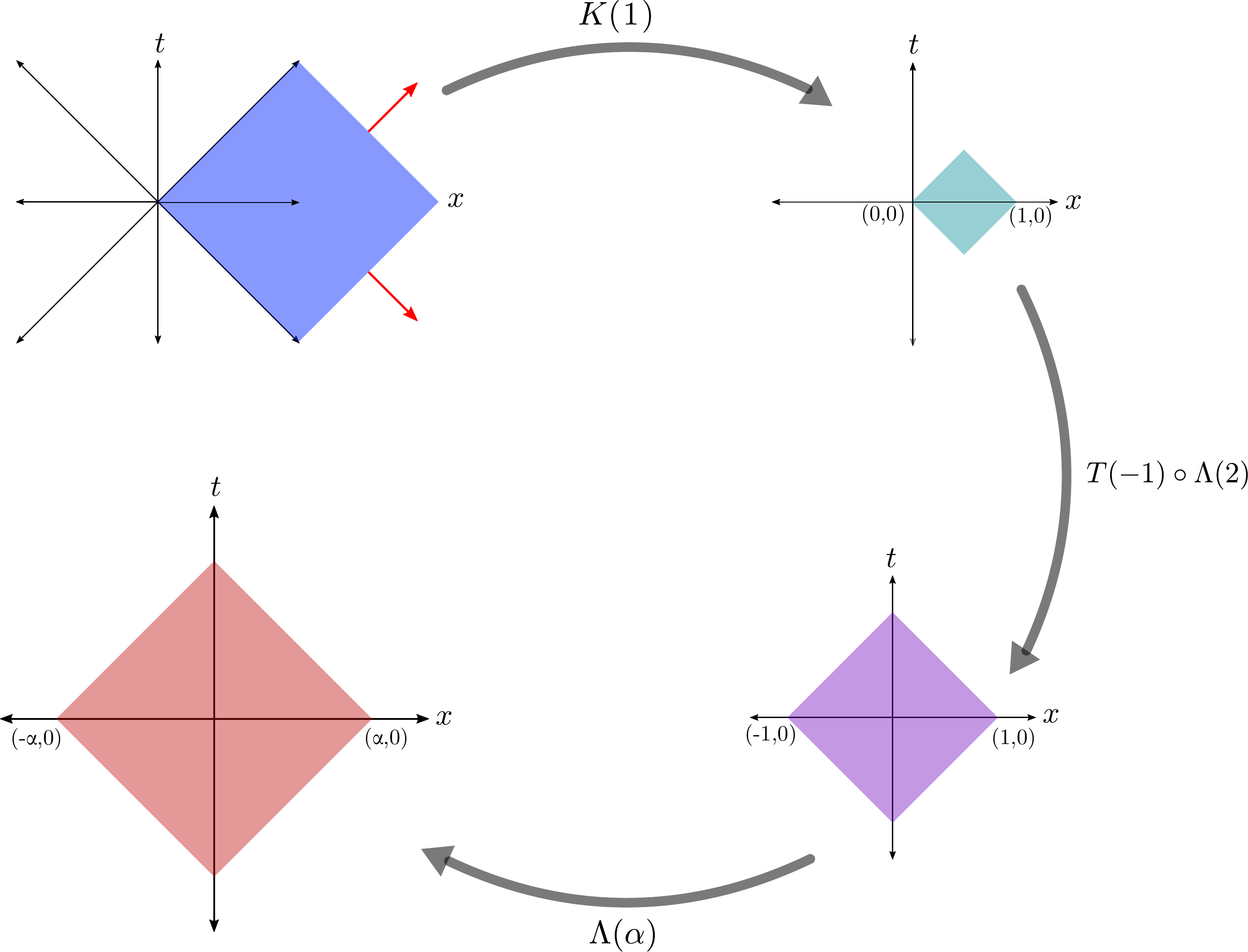}
    \caption{This figure illustrates the composite map Eq.~(\ref{eq:map-rovelli}) that yields the transformation obtained in Refs.~\cite{martinetti-1,martinetti-2}. }
    \label{fig:Rindler-to-diamond}
\end{figure}

For comparison purposes, in Refs.~\cite{martinetti-1,martinetti-2} the authors obtained the conformal mapping between $R$ and $D$ using a composition of scaling, relativistic inversion, translation, and reflection. However, the transformations used in Refs.~\cite{martinetti-1,martinetti-2} can be reconstructed by using the building blocks described in Sec.~\ref{subsec:general-conformal}. In effect, the composite map we need is
\begin{equation}
    \Lambda(\alpha) \circ T(-1) \circ \Lambda(2) \circ K(1)\;. \label{eq:map-rovelli}
\end{equation}
One can easily verify that the transformation acting on coordinates $(x_{\!_R},t_{\!_R})$ yields
\begin{align}
    t_{\!_D} &= \frac{2t_{\!_R}\alpha}{1+2x_{\!_R}+(x_{\!_R}^2-t_{\!_R}^2)} \\
    x_{\!_D} &= \frac{ \left[-1+ (x_{\!_R}^2-t_{\!_R}^2) \right]  \alpha}{1+2x_{\!_R}+(x_{\!_R}^2-t_{\!_R}^2)}
    \; ,
\end{align}
which agrees with the transformation obtained by Martinetti and Rovelli. In this case, we see that the region $R$ is first mapped into a diamond of size $1$, and then scaling and translation are used to bring the diamond to the desired size and centered at origin (Fig.~\ref{fig:Rindler-to-diamond}). The inverse transformation can be obtained by inverting the composite map. This inverted mapping is the most common form used in the literature \cite{su-ralph-1,arzano}. However, we do not need to go into detail about the inverse map, as we are going to use Eqs.~(\ref{eq:rindler-diamond-jacobson_t})--(\ref{eq:rindler-diamond-jacobson_x}) for this article.

\subsection{Static observer with a finite lifetime}\label{subsec:static-observer}
Using Eqs.~(\ref{eq:diamond-rindler}) and (\ref{eq:rindler-diamond-jacobson_t})--(\ref{eq:rindler-diamond-jacobson_x}),
we can now define the diamond coordinates as follows
\begin{align}
    \eta &= \frac{\alpha}{2} \tanh^{-1}\left( \frac{4\alpha^2 t_{\!_D}}{4\alpha^2(x_{\!_D}+\alpha) - 2\alpha[(x_{\!_D}+\alpha)^2-t_{\!_D}^2]}\right) \label{eq:diamond-jacobson} \\
    \xi &= \frac{\alpha}{2} \ln\left( \frac{16\alpha^4 t_{\!_D}^2 - \{ 4\alpha^2(x_{\!_D}+\alpha) - 2\alpha[(x_{\!_D}+\alpha)^2-t_{\!_D}^2] \}^2}{\alpha^2[(x_{\!_D}+\alpha)^2 - t_{\!_D}^2 - 4\alpha x_{\!_D}]^2} \right) 
    \; .
\end{align}
While this transformation looks complicated, we only need the trajectory for the static observer in the diamond coordinates for our discussion. This is because, in our open quantum systems model, the system is the finite-lifetime static detector sitting at the origin $(x_{\!_D} =0)$. Using the mapping in Eq.~(\ref{eq:diamond-jacobson}), one can find the relation between the diamond proper time $\eta$ for the diamond observer and the Minkowski time $t_{\!_D}$.
\begin{equation}
    \eta = \frac{\alpha}{2}\tanh^{-1}\left(\frac{2\alpha t_{\!_D}}{\alpha^2 + t_{D}^2}\right)\;. \label{eq:time-diamond-minkwoski}
\end{equation}
So, the trajectory of a diamond observer in terms of the diamond coordinates can be written as:
\begin{subequations}
\begin{align}
    t_{\!_D} &= \alpha \tanh\left(\frac{\eta}{\alpha}\right) \;,\\
    x_{\!_D} &= 0\;.
\end{align}\label{eq:static-observer-mapping}
\end{subequations}

In this section, we have clarified the procedure of defining the diamond coordinates. We have also obtained the trajectory of the diamond observer in these coordinates, and established the relation between the diamond proper time $\eta$ and the Minkowski time $t_{\!_D}$. Our objective now is to treat the diamond observer and the scalar field as a composite system. This requires specific details about the physics of the diamond observer, which we model as a finite-lifetime two-level system interacting with a scalar field. This is the subject of the next section.

\section{Unruh-DeWitt detector with scaled energy}\label{sec:udw-detector-scaled}
We can now define a finite-lifetime two-level detector as the diamond observer which interacts with a scalar field. A similar construction was used previously by Unruh and DeWitt for an accelerated detector in the region $R$ to find its thermal excitation probability \cite{birrell-davies}. These detectors are therefore called Unruh-DeWitt detectors. We are going to use a similar two-level detector as the diamond observer, but with a slight modification. In this case, the Hamiltonian of the two-level system is scaled by a factor \cite{su-ralph-1,olson-ralph} such that the energy gap between the two levels is fixed with respect to the proper time of the detector, which is the diamond time $\eta$. For this purpose, we define the Hamiltonian of the two-level system as
\begin{equation}
    H_0' = \frac{H_0}{1-\frac{t^2}{\alpha^2}}
    \label{eq:H_0'}
    \; ,
\end{equation}
 where 
\begin{equation}
 H_0 = \frac{\omega_{0}}{2} \sigma_z
     \label{eq:H_0}
 \end{equation}
  is the usual Hamiltonian of a two-level system, with energy gap $\omega_{0}$. A remark about notation is in order. 
  The time $t$ appearing in the scale factor in Eq.~(\ref{eq:H_0'}) is strictly the Minkowski time $t_D$, but we will use $t$ instead of $t_D$ for the sake of simplicity. This is mainly because $t_D$ is still the Minkowski time, albeit restricted to $[-\alpha,\alpha]$, and we are not going to use $t_R$ in the remainder of this paper---thus eliminating the source of any confusion. \\
Now we introduce the interaction Hamiltonian $H_I'$, which is usually taken to be a monopole or dipole interaction with the scalar field. In this case, to keep things simple, we will only consider the monopole interaction. The interaction Hamiltonian is linear in both the field and detector operator,
\begin{equation}
    H_I' = m\sigma_x\otimes\Phi\;.
    \label{eq:interaction-Hamiltonian}
\end{equation}
Here, $m$ stands for the coupling constant; and we take the monopole moment operator to be the Pauli matrix $\sigma_x$, which can be written as the addition of the lowering and raising operators flipping the state of the two-state system. The scalar field $\Phi$ can be decomposed into field modes satisfying the Klein-Gordon equation. In general,
\begin{equation}
    \Phi(x,t) = \sum_k  \left[ u_k(x,t) \, a_k + u^*_k(x,t) \, a_k^\dagger \right]
\; ,
\end{equation}
where $u_k$ are the field modes obtained by solving the wave equation in a particular spacetime background. It now can be seen that the state $\ket{\Psi}$ of the detector evolves according to the Schr\"odinger equation
\begin{equation}
    \frac{d\ket{\Psi}}{dt} =\left[  \frac{H_0}{1-\frac{t^2}{\alpha^2}} + H_I' \right] \ket{\Psi}   
    \; ,
    \end{equation}
in the Minkowski time $t$. However, for the detector, the proper time is the diamond time $\eta$. Thus, in its own reference frame, the Schr\"odinger equation takes the form
\begin{equation}
    \frac{d\ket{\Psi}}{d\eta} =  \bigl( H_0 + H_I \bigr) \ket{\Psi}   
    \; ,
\end{equation}
where 
\begin{equation}
H_I = \frac{H_I'}{\cosh^2(\eta/\alpha)}
\end{equation}
 is the new scaled interaction Hamiltonian. We now see the effect of the scale factor introduced in the original Hamiltonian $H_0'$: it reduces to the Hamiltonian of a two-level system in the reference frame of the diamond observer. We will now use this modified interaction Hamiltonian to find the evolution of the finite-lifetime detector. 

\section{Lindblad equation: open quantum systems approach}\label{sec:lindblad}
The state of the detector can be described by the density matrix of the two-level system. However, the system (the detector) here is not isolated as it interacts with the scalar field in a flat spacetime. Due to the interaction, we can model the evolution of the detector's state under the paradigm of open quantum systems. We can treat the field configuration here as the environment. The field is assumed to be in the Minkowski vacuum which is the natural choice of vacuum in this case as the background geometry is flat. We consider the system-environment interaction to be weak, so that the Markovian approximation holds. Under the Markovian approximation, the evolution of the reduced density matrix of the system ($\rs$) is governed by the Lindblad equation \cite{breuer-book}
\begin{equation}
    \frac{d\rho_{\!_S}}{d\eta} = -i[H_{\rm eff},\rho_{\!_S}(\eta)] + \mathcal{L}[\rs(\eta)]
    \label{eq:Lindblad}
    \;.
\end{equation}
Here, we describe the evolution in terms of the diamond proper time $\eta$. 

The right-hand side of the Lindblad equation~(\ref{eq:Lindblad}) consists of two distinctly different terms: the unitary evolution with respect to a effective Hamiltonian $H_{\rm eff}$ and the dissipative part $\mathcal{L}[\rs(\eta)]$, which is known as the Lindbladian operator. The effective Hamiltonian is of the form
\begin{equation}
    H_{\rm eff} = \frac{\Omega}{2}\sigma_z
    \label{eq:H_eff}
    \; ,
\end{equation}
where $\Omega$ is a renormalized shifted frequency that has the same origin as the Lamb shift \cite{benatti,breuer-book}. In subsequent discussions, we do not need the explicit expression for $\Omega$. The Lindbladian operator $\mathcal{L}[\rs(\eta)]$ is defined for a general interaction of the form $H_I = \sum_{\alpha} S_\alpha \otimes B_\alpha$, where $S_\alpha$ are operators in the Hilbert space of the system and $B_{\alpha}$ are operators in the Hilbert space of the environment (bath). Then, $\mathcal{L}[\rs(\eta)]$ can be written the following form,
\begin{equation}
        \mathcal{L}[\rs(\eta)]  = \sum_{\omega,\alpha,\beta} \gamma_{\alpha\beta}(\omega) \left(S_\beta(\omega) \rs S_\alpha^\dagger(\omega) - \frac{1}{2}\left\{S_\alpha^\dagger(\omega) S_\beta(\omega),\rs\right\} \right)\;,\label{eq:Lindblad-gen}
\end{equation}
where $\{ , \}$ stands for the anticommutator and the factors $\gamma_{\alpha \beta} (\omega)$ encode the contribution of the environment. The reduced density matrix $\rs$ of the system can be derived from the system-environment density matrix by averaging over the degrees of freedom of the field. This averaging leads to the factors 
$\gamma_{\alpha \beta} (\omega)$.

In our case, the interaction Hamiltonian only has one term: $\displaystyle H_I = m\sigma_x \otimes \frac{\Phi}{\cosh^2(\eta/\alpha)}$. Therefore, the double sums over $\alpha$ and $\beta$ are superfluous in the expression of the Lindblad operator, and Eq.~(\ref{eq:Lindblad-gen}) simplifies to
\begin{equation}
    \mathcal{L}[\rs(\eta)] = \sum_{\omega} \gamma(\omega) \left(S(\omega) \rs S^\dagger(\omega) - \frac{1}{2}\left\{S^\dagger(\omega) S(\omega),\rs\right\} \right)\;. \label{eq:Lindblad-nosum}
\end{equation}
Now, the field is in the Minkowski vacuum, so that its density matrix is $\rho_{\!_B} = \ket{0_M}\bra{0_M}$. In addition, $\gamma(\omega)$ is the Fourier transform 
\begin{equation}
    \gamma(\omega) = \int_{-\infty}^\infty d\Delta \eta\; e^{i\omega \Delta \eta}\; \bra{0_M}B(\eta)B(\eta')\ket{0_M}
    \label{eq:gamma-Fourier}
\end{equation}
of the two-point correlation function
\begin{equation}
 G^{+} (\Delta \eta )
 \equiv
 \bra{0_M}B(\eta)B(\eta')\ket{0_M}
 = 
 \bra{0_M}\frac{\Phi(\eta)\Phi(\eta')}{\cosh^2(\eta/\alpha)\cosh^2{(\eta'/\alpha)}}\ket{0_M}
\; ,
\label{eq:2-point-correlation}
\end{equation}
 which has an argument $\Delta \eta= \eta - \eta'$
 that only depends on the difference of the values of the two proper times involved in the Wightman function.  This effectively amounts to using $\eta$ as the argument of $G^{+}$ and setting $\eta'=0$ without any loss of generality.
This symmetry property will be verified in the next section.

In order to find the solutions to the Lindblad equation, we still need to define the factors $S_\alpha(\omega)$. For the case being analyzed, and omitting the subscripts, $S(\omega)$ is defined using projection operators $\Pi(\epsilon) = \ket{\epsilon}\bra{\epsilon}$, where $\{\ket{\epsilon}\}$ defines an eigenbasis of the Hamiltonian $H_0$ of the system; thus,
\begin{equation}
    S(\omega) = \sum_{\epsilon' - \epsilon = \omega} \Pi(\epsilon)\;S\;\Pi(\epsilon')\;.
      \label{eq:S-operator-frequency}
\end{equation}
 For the two-level system, the two energy eigenvalues are $\epsilon = \pm \omega_{0}/2$ and the eigenbasis of $H_0$ consists of the eigenvectors $\ket{\pm}_{z}$ of the $\sigma_z$ operator; with this notation, $\ket{\epsilon}= \ket{\sgn (\epsilon)}_{z}$, but we will use the standard shorthand $\ket{\pm}$ instead of $\ket{\pm}_{z}$ or $\ket{\epsilon}$. In addition, when writing the frequency components $S(\omega)$, one considers all possible transitions in the two-level system, according to Eq.~(\ref{eq:S-operator-frequency}), i.e., $\omega = \epsilon' - \epsilon = \pm \omega_{0}, 0$.
Moreover, for the interaction Hamiltonian~(\ref{eq:interaction-Hamiltonian}), the operator $S$ is just $\sigma_x$. (For a generalization of this choice, see Appendix~\ref{sec:comparative-framework-Bloch}.) Therefore, the only three relevant operators $S(\omega)$ are given by
\begin{align}
    S(+\omega_{0}) &= \ket{-}\bra{-}\sigma_x \ket{+}\bra{+} = \ket{-}\bra{+}     \label{eq:S-operator_plus}\\
    S(-\omega_{0}) &= \ket{+}\bra{+}\sigma_x \ket{-}\bra{-} = \ket{+}\bra{-}     \label{eq:S-operator_minus}\\
    S(0) &= 0     \label{eq:S-operator_zero}
    \;.
\end{align}

In short, we now have all the ingredients to solve the Lindblad equation. One method for its solution, which is ideally tailored for a two-level system, consists in representing the reduced density matrix $\rho_{S}(\eta)$ as a 
Bloch vector $\ket{\rho (\eta)} $ with components $\rho_{i}(\eta)$ ($i=1,2,3$) defined 
via the generic expansion for operators in terms of the basis of the three Pauli matrices and the unit matrix,
\begin{equation}
    \rs = \frac{1}{2}\left(\mathit{I} + \sum_{i=1}^{3}\rho_i\sigma_i \right) \label{eq:density-decomp}
    \; .
\end{equation}
The prefactor of one half guarantees the unit-trace condition of the density matrix, and otherwise defines conventionally the components $\rho_{i}(\eta)$. The equation for the components $\rho_{i}(\eta)$ of the Bloch vector can be found by expanding all the operators from Eqs.~(\ref{eq:Lindblad}), (\ref{eq:H_eff}), and (\ref{eq:density-decomp}) in this basis.
Then, the Lindblad equation reduces a Schr\"{o}dinger-like vector equation for the three-component Bloch vector 
$\ket{\rho}$ of the form
\begin{equation}
    \frac{d\ket{\rho}}{d\eta}  = 
    - 2 \mathcal{H}\ket{\rho} + \ket{n} 
    \label{eq:linear-system-rho}
    \; ,
    \end{equation}
    where the operator $\mathcal{H}$ has the 
      $3 \times 3$ matrix representation
  \begin{equation}
  \mathcal{H}
  =
    \begin{pmatrix}
    A/2 & \Omega/2 & 0\\
   - \Omega/2 & A/2 & 0\\
    0 & 0 & A
    \end{pmatrix}
        \label{eq:mathcal-H-matrix}
        \; ,
\end{equation}
and $ \ket{n} $ is the constant vector $(0,0, -2B)$. 
Thus, in matrix form,
\begin{equation}
   \begin{pmatrix}
    \dot{\rho}_1\\
    \dot{\rho}_2\\
    \dot{\rho}_3
    \end{pmatrix}
    = - 2  \mathcal{H}
     \begin{pmatrix}
    \rho_1\\
    \rho_2\\
    \rho_3
    \end{pmatrix} + \begin{pmatrix}
    0 \\
    0\\
    -2B
    \end{pmatrix} 
    \; .
    \label{eq:linear-system-rho_matrix}
\end{equation}
In these equations,
\begin{equation}
 A = \frac{1}{2}  [\gamma(\omega_{0})+\gamma(-\omega_{0})]
\; \; ; \; \;
B  = \frac{1}{2}[\gamma(\omega_{0})-\gamma(-\omega_{0})]\;.
\label{eqA-B-C-values}
\end{equation}

 In the next section, we are going to find the system's steady-state density matrix and investigate its properties, which determine the thermal character of the system.

\section{Steady-state reduced density matrix and diamond temperature}\label{sec:steady-state}
In the evolution of the system interacting with the environment, 
a steady state is achieved asymptotically as $\eta \rightarrow \infty$. At the steady state, the left-hand side of Eq.~(\ref{eq:linear-system-rho}) becomes zero. This leads to the equilibrium values of the coefficients $\rho_i$ which we denote $\rho_{i}(\infty)$; the corresponding Bloch vector $\ket{\rho_{S}}_{\!_\infty}$ is
\begin{equation}
    \ket{\rho_{\!_S}}_\infty =\frac{1}{2} \mathcal{H}^{-1} \ket{n} = -
        \frac{B}{A} \begin{pmatrix}
    0\\
    0\\
1
    \end{pmatrix}
    \; .
    \label{eq:steady-state-rho}
\end{equation}
Then, there is only one nonzero component of the density matrix in the Bloch representation (corresponding to the $z$ axis), which is due to the form of the initial Hamiltonian $H_0$ with a similar decomposition. 

Moreover, one can further analyze the evolution of the density matrix by explicitly solving the linear equation~(\ref{eq:linear-system-rho}) by direct integration,
\begin{equation}
\ket{\rho (\eta)}
=
e^{-2 \mathcal{H} \eta} \,    \ket{\rho (0)} 
+
(1-e^{-2 \mathcal{H} \eta }) \,   \ket{\rho_{\!_\infty}} 
    \; ,
\end{equation}
which verifies its asymptotic approach to the form~(\ref{eq:steady-state-rho}).
This analysis can be further extended by evaluating the exponential $e^{-2 \mathcal{H} \eta }$ of the matrix~(\ref{eq:mathcal-H-matrix}) of degree three by using the Cayley-Hamilton theorem, which displays the nonunitary dissipation and decoherence behavior governed by the thermal Boltzmann factor~\cite{benatti,yu-zhang-1}. The coefficients that govern both the steady-state density matrix and its evolution with dissipation and decoherence, can be found by evaluating $\gamma(\omega)$, as can be seen from Eq.~(\ref{eqA-B-C-values}); this is given by the Fourier transform of the two-point correlation function, as in Eqs.~(\ref{eq:gamma-Fourier}) and  (\ref{eq:2-point-correlation}).
The  Wightman function $G^{+}$ only depends on the difference of the values of the two proper times, i.e., it is function 
of the simple argument $\Delta \eta= \eta - \eta'$. This is explicitly verified below. In flat spacetime, the two-point correlation function 
$\bra{0_M}\Phi({\bf x},t)\Phi({\bf x}',t')\ket{0_M}$ has the expression
\begin{equation}
  \bra{0_M}\Phi({\bf x},t)\Phi({\bf x}',t')\ket{0_M} = -\frac{1}{4\pi^2} \frac{1}{|t-t'-i\epsilon|^2 - |{\bf x}-{\bf x}'|^2}\;.  
  \label{eq:2point-Minkowski}
\end{equation}
For a finite-lifetime diamond detector, the trajectory of the detector is defined by the relation between Minkowski time $t$ and diamond time $\eta$, as the detector is located at the origin (${\bf x} = 0$) for the entirety of its existence. In Eq.~(\ref{eq:static-observer-mapping}), we found that the two times are related by 
$t = \alpha \tanh\left(\eta/\alpha\right)$. Replacing this in Eq.~(\ref{eq:2point-Minkowski}), we get
\begin{equation}
G^{+} (\eta - \eta') =
    \bra{0_M} \frac{\Phi(\eta)\Phi(\eta')}{\cosh^2(\eta/\alpha) \cosh^2(\eta'/\alpha)}
    \ket{0_M} = -\frac{1}{4\pi^2 \alpha^2}\,  \frac{1}{\sinh^2\left[ (\eta-\eta')/\alpha - i \epsilon \right] }
    \; .
\end{equation}
where we have absorbed a positive function of $\eta,\,\eta',\,\alpha$ into $\epsilon$. This verifies, as expected, that the scaled two-point function above is dependent only on the proper time interval $\eta-\eta'$. The value of $\gamma(\omega)$ can be computed from
the Fourier transform~(\ref{eq:gamma-Fourier}) 
\begin{equation}
    \gamma(\omega) = -\frac{1}{4\pi^2 \alpha^2} \int_{-\infty}^\infty d\eta \; e^{i\omega \eta}\;
    \frac{1} { \sinh^{2} \left(\eta/\alpha - i \epsilon \right)}  = \frac{\omega}{2\pi} \frac{1}{1-e^{-\pi\alpha\omega}} \label{eq:gamma-omega}
    \; .
\end{equation}
This important integral is the key that yields a thermal density matrix, as shown below; its evaluation requires the $i\epsilon $ prescription that places the pole corresponding to $\eta=0$ in the upper half plane, as is easily shown by contour integration: see Appendix~\ref{app:integral-wightman} for the evaluation of the integral.

Using Eq.~(\ref{eq:gamma-omega}), the only non-zero coefficient in the expansion of the density matrix in Eq.~(\ref{eq:density-decomp}) reduces to
\begin{equation}
    \rho_{\!_3}(\infty) = 
        - \frac{B}{A} = \frac{\gamma(-\omega_{0})-\gamma(\omega_{0})}{\gamma(\omega_{0})+\gamma(-\omega_{0})} = \frac{1-e^{\pi\alpha\omega_{0}}}{1+e^{\pi\alpha\omega_{0}}}
\; .
\end{equation}
The steady-state density matrix then can be written using the expansion of Eq.~(\ref{eq:density-decomp}),
\begin{equation}
    \rs(\infty) = \begin{pmatrix}
 \displaystyle 
    \frac{1}{1+e^{\pi\alpha\omega_{0}}} & 0\\
    0 &  \displaystyle  \frac{e^{\pi\alpha\omega_{0}}}{1+e^{\pi\alpha\omega_{0}}} 
    \end{pmatrix} = \frac{e^{-\pi\alpha H_0}}{{\rm Tr}\left[e^{-\pi\alpha H_0}\right]}
    \; .
\end{equation}
This expression has the same form as the density matrix of the two-level system kept in a thermal bath with temperature $T = 1/\beta$,
\begin{equation}
    \rs =  \frac{e^{-\beta H_0}}{{\rm Tr}\left[e^{-\beta H_0}\right]}
    \; ,
\end{equation}
if one identifies the temperature with the quantity $T = 1/(\pi\alpha)$. 
Clearly, the steady-state density operator of the diamond observer has the form of a thermal density matrix.
Thus, the diamond observer experiences a thermal bath in the Minkowski vacuum with temperature inversely proportional to its lifetime,
\begin{equation}
T_D = \frac{1}{\pi \alpha} = \frac{2}{\pi \mathcal{T}} 
\; .   
\label{eq:diamond-temperature}
\end{equation}
In conclusion, Eq.~(\ref{eq:diamond-temperature}) gives the diamond temperature $T_D$ \cite{su-ralph-1} in terms of 
 $\mathcal{T} = 2\alpha$, the lifetime of the particle or the size of the diamond (according to Minkowski coordinates).
 
 \section{Conclusions and outlook}\label{sec:conclusions}
In this article, we have introduced a systematic, rigorous method for defining the diamond coordinates, which extends existing work in the literature. Furthermore, we have used the Lindblad equation to identify the state of the diamond observer in an open quantum systems approach. This approach provides another meaning for the temperature of the diamond: it is the temperature of the thermal steady-state reduced density matrix of a two-state finite-lifetime detector with a scaled energy gap. Such methodology has obvious advantages, as the reduced density matrix provides a full characterization of the state of the system. Moreover, the open quantum systems approach opens up new directions for future work. First, it leads to the quantum information theoretic description of the system where one can cast the Lindblad equation in terms of Kraus operators \cite{nielsen-chuang,kraus-lindblad}. Second, it suggests the feasibility of simulating the system-environment interaction in the diamond in a quantum computer. Third, it provides a methodology for modeling the evolution of the correlation between two-level finite-lifetime detectors, possibly leading to an entanglement harvesting protocol. Finally, this framework has great potential for revisiting a broad range of problems at the intersection of general relativity and quantum field theory.

\acknowledgments{}

H.E.C. acknowledges support by the University of San Francisco Faculty Development Fund. This material is based upon work supported by the Air Force Office of Scientific Research under Grant No. FA9550-21-1-0017 (C. R. O. and A.C.).

\begin{appendix}

\section{Evaluation of integral in Eq.~(\ref{eq:gamma-omega})}\label{app:integral-wightman}
\begin{figure}
    \centering
    \includegraphics[width=0.5\linewidth]{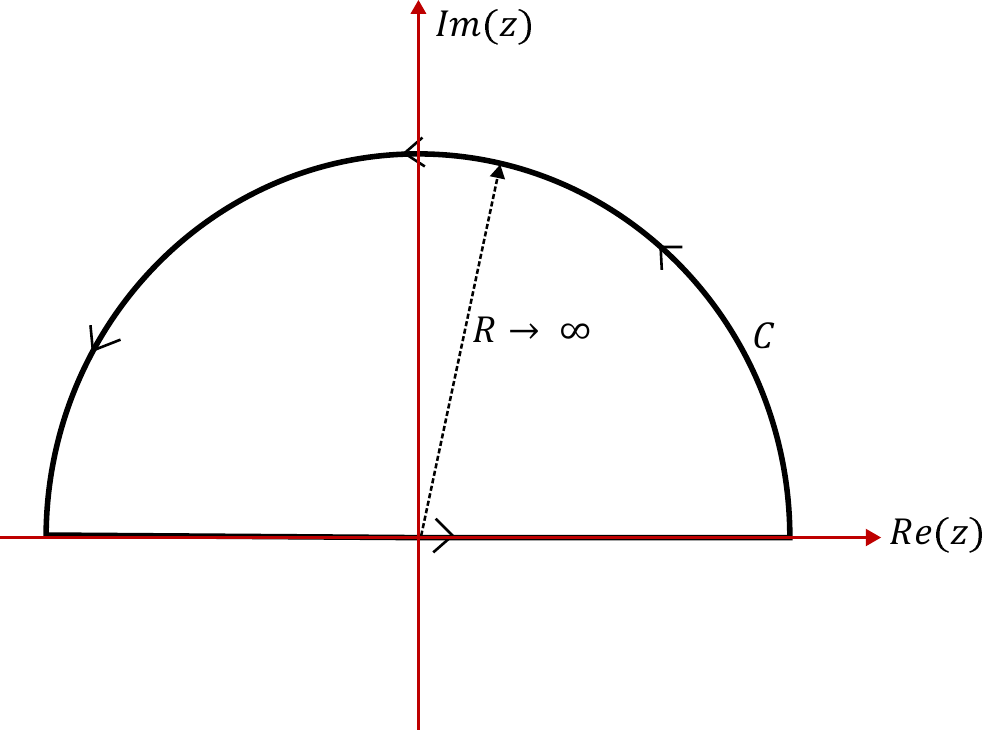}
    \caption{The semicircular contour C, used to evaluate Eq.~(\ref{eq:transformed-integral}), with radius $R\rightarrow\infty$, closed in the upper-half plane.}
    \label{fig:contour-integral}
\end{figure}

In this appendix, we show the steps to evaluate the Fourier transform of the Wightman function given in Eq.~(\ref{eq:gamma-omega}):
\begin{equation}
    \mathcal{I} = \int_{-\infty}^{\infty} d\eta\;\;  \frac{e^{i\omega\eta}}{\sinh^2(\eta/\alpha - i \epsilon)} \;.\label{eq:integral-to-find}
\end{equation}
We start by a redefinition of variables: $z=\eta/\alpha$ and $q=\omega\alpha$. So, we can now recast Eq.~(\ref{eq:integral-to-find}) in the following way:
\begin{equation}
    \hat{\mathcal{I}} \equiv \frac{\mathcal{I}}{\alpha} =  \int_{-\infty}^{\infty} dz\;\;\frac{e^{iqz}}{\sinh^2(z - i \epsilon)} \equiv  \int_{-\infty}^{\infty} dz\;\;e^{iqz} f(z)\;. \label{eq:transformed-integral}
\end{equation}
The integral $\hat{\mathcal{I}}$ can now be evaluated using complex contour integration. Using Jordan's Lemma, we can close the contour in the upper-half complex z-plane since $q>0$,
\begin{equation}
    \int_{-\infty}^{\infty} dz\;\;e^{iqz} f(z) = \oint_C\, dz\;\; e^{iqz} f(z) = 2\pi i \sum_k {\rm Res}\,(z_k)\;, \label{eq:Jordans-lemma}
\end{equation}
where $C$ is the semicircular contour with infinite radius closed in the upper-half $z$ plane (see Fig.~\ref{fig:contour-integral}). Res\,$(z_k)$ is the residue of the function $e^{iqz} f(z)$ for the pole at $z=z_k$.

So, to evaluate the integral in Eq.~(\ref{eq:transformed-integral}), we need to find the poles of the integrand in the upper-half z-plane, and calculate their corresponding residues. The integrand has second-order poles at the zeros of the function $\sinh^2(z-i\epsilon)$, i.e.,
\begin{equation}
    \sinh^2(z_k-i\epsilon) = 0\qquad \implies\qquad z_k = i\pi k + i\epsilon\;,
\end{equation}
where $k\in \mathbb{Z}$ (all integers). Only the poles with integers $k \geq 0$ are enclosed by the contour. The
residues are
\begin{equation}
    {\rm Res}\,(z_k) = iqe^{-k\pi q}\;.
\end{equation}

Using Eq.~(\ref{eq:Jordans-lemma}), we can now evaluate the integral in Eq.~(\ref{eq:transformed-integral}):
\begin{equation}
    \hat{\mathcal{I}} = 2\pi i (iq) \sum_{k=0}^\infty e^{-k\pi q} = -\frac{2\pi q}{1-e^{-\pi q}}\;.
\end{equation}
Restoring the original variables back, we get
\begin{equation}
    \mathcal{I} = -2\pi\omega\alpha^2(1-e^{\pi\omega\alpha})^{-1}\;,
\end{equation}
which gives us the final result of Eq.~(\ref{eq:gamma-omega}).

\section{Generalized comparative framework for the Bloch-Vector Form of the Lindblad Equation}
\label{sec:comparative-framework-Bloch}

The calculation shown in the main text, Sec.~\ref{sec:udw-detector-scaled},
leading to the Schr\"{o}dinger-like Bloch-vector form of the Lindblad equation~(\ref{eq:linear-system-rho}),
specializes to an interaction Hamiltonian~(\ref{eq:interaction-Hamiltonian}).
This Hamiltonian involves a system's operator that is proportional to $\sigma_{x}$ alone, i.e. $S= \sigma_{x}$;
or, more generally, $S= m \sigma_{x}$, including a coupling strength $m$.
In this Appendix, we present a larger framework that is a modified (and simplified) alternative
 to the one in Ref.~\cite{benatti}, adapted for a comparison of different literature results using a uniform notation.
This should help clarify some possible confusion regarding the selection of numerical factors
in the final equations for $\mathcal{H}$ and $[ \ket{n} ] $ as needed for Eq.~(\ref{eq:linear-system-rho}), 
including those found in published references---see Eqs.~(\ref{eq:linear-system-rho_matrix_general}) and (\ref{eq:mathcal-H-matrix_general}).
Additional generalizations are possible, following the techniques of Ref.~\cite{benatti}---but unlike this reference,
we are only considering the case where the effective Hamiltonian~(\ref{eq:H_eff}) of the two-level system 
has a quantization axis along $z$. 
        
The generic system's operators can be written in the form
\begin{equation}
S= \sum_{\mu} m_{\mu} \, \sigma_{\mu}
\; ,
\end{equation}
with a set of coupling strengths $m_{\mu}$ ($\mu = 0,1,2,3$), and
 with the usual conventions for the Pauli matrices:
$\sigma_{j} \equiv \sigma_{x} , \sigma_{y}, \sigma_{z}$, for $j=1,2,3$,
and $\sigma_{0} \equiv \mathit{I}$ for the identity matrix.
In this basis, the general resolution of the density matrix is given by Eq.~(\ref{eq:density-decomp}).
As discussed in the explanation for Eqs.~(\ref{eq:S-operator-frequency})--(\ref{eq:S-operator_zero}),
there are only three possible transitions in the two-level system, with $\omega = \epsilon' - \epsilon = \pm \omega, 0$,
which give the only three relevant components of $S(\omega)$:
\begin{equation}
S_{\pm} \equiv S( \pm \omega_{0}) = \left( m_{1} \pm i m_{2} \right) \, P_{\mp \pm} \; \; , \; \;  \; \; \;
S_{0} \equiv S(0) = m_{0}       \mathit{I} + m_{3} \sigma_{3}
\; ,
\end{equation}
with the operators $ P_{\mp \pm} = \ket{\mp}\bra{\pm}=  \left( \sigma_{x} \mp i \sigma_{y} \right)\!/2$.

 For the field operators, we will keep the parameters defined in
 Eq.~(\ref{eqA-B-C-values}), and further expand them to include the transition of zero frequency, i.e.,
 \begin{equation}
A = \frac{1}{2}  [\gamma(\omega_{0})+\gamma(-\omega_{0})]
\; \; ; \; \;
B  = \frac{1}{2}[\gamma(\omega_{0})-\gamma(-\omega_{0})]
\; \; ; \; \;
C = \gamma(0) - A
\; .
\label{eqA-B-C-values_general}
\end{equation}

   The intermediate steps in the calculation of the components of 
 Eq.~(\ref{eq:linear-system-rho}) are as follows. They involve
 repeated application of the definitions of the Pauli matrices and their product relations
 $\sigma_{j} \sigma_{k} = \delta_{jk} \mathit{I} + i \epsilon_{jkl} \sigma_{l}$ (as well as use of the Einstein summation convention for any pair of repeated indices).
 First,
 for the products in the first term of Eq.~(\ref{eq:Lindblad-nosum}), at each transition frequency,
 \begin{align}
& S_{\pm} \, \rho \, S_{\pm}^{\dagger} 
=
 \frac{1}{2}  \left( m_{1}^{2}  + m_{2}^{2} \right) 
  \left( 1 \pm \rho_{3} \right) 
 \Pi_{\mp} 
 \\
&   S_{0} \, \rho  \, S_{0}^{\dagger} 
=
\left[ \frac{1}{2}   \left(m_{0}^{2}  + m_{3}^{2} \right) + m_{0} m_{3} \rho_{3} \right] \mathit{I}
+\left[  m_{0} m_{3}
+  \frac{1}{2}   \left(m_{0}^{2}  + m_{3}^{2} \right) \rho_{3} \right] \sigma_{3} 
\nonumber
\\
& \qquad \qquad \qquad \qquad  \qquad \qquad \qquad \qquad  \qquad
 +  \frac{1}{2}   \left(m_{0}^{2}  - m_{3}^{2} \right) \,  
\left( \rho_{1}  \sigma_{1} +  \rho_{2}  \sigma_{2} \right) 
\; ,
\end{align}
where $ \Pi_{\pm} $ denotes the projectors $ \Pi_{\pm} =  \left( \mathit{I} \pm \sigma_{3} \right)/2$.
Second, for the anticommutators in the second term of Eq.~(\ref{eq:Lindblad-nosum}),
the products 
$S_{\pm}^{\dagger}  S_{\pm}
= \left( m_{1}^{2}  + m_{1}^{2} \right) \, \Pi_{\pm} $ and
$S_{0}^{\dagger} S_{0}
=  \left(m_{0}^{2}  + m_{3}^{2} \right) \,  \mathit{I} 
+ 2m_{0} m_{3} \sigma_{3}$
give
 \begin{align}
& \frac{1}{2} \,
\left\{
S_{\pm}^{\dagger}   S_{\pm} , \rho
\right\}
=
\left( m_{1}^{2}  + m_{1}^{2} \right)
\,
\left( \frac{1}{2} \Pi_{\pm} 
\pm \frac{1}{4} \rho_{3} \mathit{I}
+ \frac{1}{4} \rho_{j} \sigma_{j} 
\right)
\\
& \frac{1}{2} \,
\left\{
S_{0}^{\dagger} S_{0} , \rho
\right\}
=
\frac{1}{2} S_{0}^{\dagger} S_{0}
+ m_{0} m_{3} \rho_{3} \mathit{I}
+ \frac{1}{2}   \left(m_{0}^{2}  + m_{3}^{2} \right) \,  
\left( \rho_{1}  \sigma_{1} +  \rho_{2}  \sigma_{2} +  \rho_{3}  \sigma_{3}
\right)
\; .
\end{align}
Then, the building blocks of the Lindbladian include
\begin{equation}
\begin{aligned}
&
L_{\pm} 
= S_{\pm} \rho S_{\pm}^{\dagger}
 - \frac{1}{2} \left\{ S_{\pm}^{\dagger}  S_{\pm} , \rho \right\}
=
\left( m_{1}^{2}  + m_{2}^{2} \right) \, 
\left[
\mp \frac{1}{2} \sigma_{3}
 - \frac{1}{4} \left(  \rho_{1} \sigma_{1} + \rho_{2} \sigma_{2} + 2 \rho_{3} \sigma_{3} \right) 
 \right]
\\
& 
L_{0} 
= S_{0} \rho S_{0}^{\dagger}
 - \frac{1}{2} \left\{ S_{0}^{\dagger}  S_{0} , \rho \right\}
= - m_{3}^{2} 
\left( \rho_{1}  \sigma_{1} +  \rho_{2}  \sigma_{2}
\right)
\; ,
\end{aligned}
\end{equation}
so that
\begin{align}
&{\mathcal L}
= \gamma_{+} L_{+}  +  \gamma_{-} L_{-}  +  \gamma_{0} L_{0} 
\\
&
= \left( m_{1}^{2}  + m_{2}^{2} \right) \, 
\left[
 \frac{1}{2} \sigma_{3} \left(-\gamma_{+}+ \gamma_{-} \right)
 - \frac{1}{4} \left(  \rho_{1} \sigma_{1} + \rho_{2} \sigma_{2} + 2 \rho_{3} \sigma_{3} \right) 
 \left(\gamma_{+}+ \gamma_{-} \right)
\right]
-
\gamma_{0} m_{3}^{2} 
\left( \rho_{1}  \sigma_{1} +  \rho_{2}  \sigma_{2}
\right)
\; ,
\label{eq:Lindbladian-1}
\end{align}
with the notation $\gamma_{\pm} = \gamma (\pm \omega_{0})$ and
$\gamma_{0} = \gamma (0)$. 
Rearranging the terms in Eq.~(\ref{eq:Lindbladian-1}),
\begin{equation}
{\mathcal L}
= 
-B' \sigma_{3} - \frac{1}{2} A' 
\left(  \rho_{1} \sigma_{1} + \rho_{2} \sigma_{2} + 2 \rho_{3} \sigma_{3} \right)
- \gamma_{0} m_{3}^{2} \left(  \rho_{1} \sigma_{1} + \rho_{2} \sigma_{2} \right)
\; ,
\label{eq:Lindbladian-2}
\end{equation}
where $A'= \left( m_{1}^{2}  + m_{2}^{2} \right) A$ 
and $B'=  \left( m_{1}^{2}  + m_{2}^{2} \right)  B$.

As a result, with the resolution in terms of Pauli operators,
the Lindblad equation~(\ref{eq:Lindblad}) takes the form
\begin{equation}
\frac{1}{2}  \dot{\rho}_{k} \sigma_{k}
=
- \frac{\Omega}{2} \,  \epsilon_{3kj} \rho_{j} \sigma_{k}
- B' \sigma_{3} 
-
\left( \frac{1}{2} A' + m_{3}^2 \gamma_{0} \right)
 \left(  \rho_{1} \sigma_{1} + \rho_{2} \sigma_{2} \right)
 - \frac{1}{2} (2 A')    \rho_{3} \sigma_{3} 
 \; ,
\label{eq:Lindblad-eq_Paulian}
\end{equation}
where the first term on the right-hand side arises from the unitary-evolution commutator
$[H_{\rm eff}, \rho] 
= \bigl[ (\Omega/2) \sigma_{3},  \left( \mathit{I} + \rho_{j} \sigma_{j} \right)/2  \bigr]
= - i (\Omega/2) \, \epsilon_{3kj} \rho_{j} \sigma_{k}$.
Also, $
\frac{1}{2} A' + m_{3}^2 \gamma_{0} = 
\left[ \frac{1}{2} \left( m_{1}^{2}  + m_{2}^{2} \right) + m_{3}^2 \right] A + m_{3}^{2}  C$.
Then, the linear combinations of Eq.~(\ref{eq:Lindblad-eq_Paulian}) in the Pauli-matrix
basis lead to the components satisfying 
the Bloch-vector form of the Lindblad equation
\begin{equation}
   \begin{pmatrix}
    \dot{\rho}_1\\
    \dot{\rho}_2\\
    \dot{\rho}_3
    \end{pmatrix}
    = - 2  \mathcal{H}
     \begin{pmatrix}
    \rho_1\\
    \rho_2\\
    \rho_3
    \end{pmatrix} + \begin{pmatrix}
    0 \\
    0\\
    -2  \left( m_{1}^{2}  + m_{2}^{2} \right) \, B
    \end{pmatrix} 
    \; ,
    \label{eq:linear-system-rho_matrix_general}
\end{equation}
where the last term, $ \ket{n} $, is the constant vector 
$\bigl( 0,0, -2  \left( m_{1}^{2}  + m_{2}^{2} \right) \, B \bigr)$,
and the 
operator $\mathcal{H}$ has the 
      $3 \times 3$ matrix representation
  \begin{equation}
  \! 
  \mathcal{H}
  =
    \begin{pmatrix}
  \!  \left[ \frac{1}{2} \left( m_{1}^{2}  + m_{2}^{2} \right) + m_{3}^2 \right] A + m_{3}^{2}  C
    & \Omega/2 & 0\\
\! \!   - \Omega/2 & 
   \left[ \frac{1}{2} \left( m_{1}^{2}  + m_{2}^{2} \right) + m_{3}^2 \right] A + m_{3}^{2}  C & 0 \\
       0 & 0 & 
   \! \!  \left( m_{1}^{2}  + m_{2}^{2} \right) A
    \end{pmatrix}
        \label{eq:mathcal-H-matrix_general}
        \, .
\end{equation}

 In conclusion, if one considers a nonzero interaction with contributions from all the Pauli matrices, 
 and with all the couplings set to the same value $m_{\mu} = m=1 $, 
 then the components of the Bloch-vector equation reduce to 
  \begin{equation}
  \mathcal{H}
  =
    \begin{pmatrix}
 2A + C    & \Omega/2 & 0 \\
   - \Omega/2 &  2 A + C  & 0 \\
       0 & 0 &   2 A
    \end{pmatrix}
      =
    \begin{pmatrix}
 A + \gamma (0)    & \Omega/2 & 0 \\
   - \Omega/2 &  A + \gamma (0)  & 0 \\
       0 & 0 &   2 A
    \end{pmatrix}
        \label{eq:mathcal-H-matrix_general-all}
        \; ,
\end{equation}
and $ \ket{n}  = (0,0,- 4B)$; this agrees with the results in the original paper of Ref.~\cite{benatti}
(for the particular case of an effective Hamiltonian~(\ref{eq:H_eff}) with quantization axis along $z$). 

But when using only the Pauli matrix $\sigma_{x}$, as in the main text of our paper, 
then Eqs.~(\ref{eq:mathcal-H-matrix}) and (\ref{eq:linear-system-rho_matrix})
are obtained.
Clearly, the latter case cannot be obtained from the former one by just enforcing the lack of transitions with zero frequency, i.e., by setting $\gamma (0) = 0$, because both Pauli matrices $\sigma_{x}$ and $\sigma_{y}$ 
generate this outcome but contribute separately to the final result 
in Eqs.~(\ref{eq:linear-system-rho_matrix_general}) and (\ref{eq:mathcal-H-matrix_general})---this is
a subtlety that can easily lead to inconsistent results
(e.g., the combination of Eqs.~(4), (13), and (15) in Ref.~\cite{yu-zhang-1}).

\end{appendix}


\begin{thebibliography}{}
\bibitem{hawking1}
S. W. Hawking, Black hole explosions?, Nature, \textbf{248(5443)}, 30-31. (1974)
\bibitem{hawking2}
S. W. Hawking, Particle creation by black holes, Communications in Mathematical Physics, \textbf{43(3)}, 199-220. (1975)
\bibitem{hawking3}
S. W. Hawking, Black holes and thermodynamics, Physical Review D, \textbf{13(2)}, 191. (1976)
\bibitem{unruh76}
W. G. Unruh, Notes on black-hole evaporation, Physical Review D, \textbf{14(4)}, 870. (1976)
\bibitem{wald75}
R. M. Wald, On particle creation by black holes, Communications in Mathematical Physics, \textbf{45(1)}, 9-34. (1975)
\bibitem{parker75}
L. Parker, Probability distribution of particles created by a black hole, Physical Review D, \textbf{12(6)}, 1519. (1975)
\bibitem{ufd76}
P. C. Davies, S. A. Fulling, and W. G. Unruh, Energy-momentum tensor near an evaporating black hole, Physical Review D, \textbf{13(10)}, 2720. (1976)
\bibitem{davies78}
P. C. Davies, Thermodynamics of black holes, Reports on Progress in Physics, \textbf{41(8)}, 1313. (1978)
\bibitem{df77}
P. C. Davies, S. A. Fulling, Radiation from moving mirrors and from black holes, Proceedings of the Royal Society of London. A. Mathematical and Physical Sciences, \textbf{356(1685)}, 237-257. (1977)
\bibitem{israel76}
W. Israel, Thermo-field dynamics of black holes, Physics Letters A, \textbf{57(2)}, 107-110. (1976)
\bibitem{carlip1}
S. Carlip, Black hole entropy from conformal field theory in any dimension, Physical Review Letters, \textbf{82(14)}, 2828. (1999)
\bibitem{carlip2}
S. Carlip, Black hole entropy from horizon conformal field theory. Nuclear Physics B-Proceedings Supplements, \textbf{88}, 10-16. (2000)
\bibitem{birminghamsen}
D. Birmingham, and S. Sen, Exact black hole entropy bound in conformal field theory, Physical Review D, \textbf{63(4)}, 047501. (2001)
\bibitem{cardy}
H. W. J. Bl\"{o}te, J. L. Cardy, and M. P. Nightingale, Conformal invariance, the central charge, and universal finite-size amplitudes at criticality, Physical Review Letters, \textbf{56(7)}, 742. (1986)
\bibitem{nhcamblong}
H. E. Camblong, and C. R. Ord\'o\~nez, Black hole thermodynamics from near-horizon conformal quantum mechanics, Physical Review D, \textbf{71(10)}, 104029. (2005)
\bibitem{nhcamblong-sc}
H. E. Camblong, and C. R. Ord\'o\~nez, Semiclassical Methods in Curved Spacetime and Black Hole Thermodynamics, Physical Review D, \textbf{71(12)}, 124040. (2005)
\bibitem{nhcamblong-tightness}
H. E. Camblong, and C. R. Ord\'o\~nez, 
Conformal tightness of holographic scaling in black hole thermodynamics, 
Classical and Quantum Gravity, \textbf{30}, 175007. (2013)
\bibitem{chakraborty1}
H. E. Camblong, A. Chakraborty, and C. R. Ord\'o\~nez, Near-horizon aspects of acceleration radiation by free fall of an atom into a black hole, Physical Review D, \textbf{102(8)}, 085010. (2020)
\bibitem{chakraborty2}
A. Azizi, H. E. Camblong, A. Chakraborty, C. R. Ord\'o\~nez, and M. O. Scully, Acceleration radiation of an atom freely falling into a Kerr black hole and near-horizon conformal quantum mechanics, Physical Review D, \textbf{104(6)}, 065006. (2021)
\bibitem{chakraborty3}
A. Azizi, H. E. Camblong, A. Chakraborty, C. R. Ord\'o\~nez, and M. O. Scully, Quantum optics meets black hole thermodynamics via conformal quantum mechanics. I. Master equation for acceleration radiation. Physical Review D, \textbf{104(8)}, 084086. (2021)
\bibitem{chakraborty4}
A. Azizi, H. E. Camblong, A. Chakraborty, C. R. Ord\'o\~nez, and M. O. Scully, Quantum optics meets black hole thermodynamics via conformal quantum mechanics: II. Thermodynamics of acceleration radiation, Physical Review D, \textbf{104(8)}, 084085. (2021)
\bibitem{martinetti-1}
P. Martinetti, and C. Rovelli, Diamond's temperature: Unruh effect for bounded trajectories and thermal time hypothesis, Classical and Quantum Gravity, \textbf{20(22)}, 4919. (2003)
\bibitem{su-ralph-1}
D. Su, and T. C. Ralph, Spacetime diamonds. Physical Review D, \textbf{93(4)}, 044023. (2016)
\bibitem{su-ralph-2}
J. Foo, S. Onoe, M. Zych, and T. C. Ralph, Generating multi-partite entanglement from the quantum vacuum with a finite-lifetime mirror,  New Journal of Physics, \textbf{22(8)}, 083075. (2020)
\bibitem{light-cone}
T. De Lorenzo, and A. Perez, Light cone thermodynamics, Physical Review D, \textbf{97(4)}, 044052. (2018)
\bibitem{jacobson}
T. Jacobson, and M. Visser, Gravitational thermodynamics of causal diamonds in (A)dS, SciPost Physics, \textbf{7(6)}. (2019)
\bibitem{benatti}
F. Benatti, and R. Floreanini, Entanglement generation in uniformly accelerating atoms: Reexamination of the Unruh effect, Physical Review A, \textbf{70(1)}, 012112. (2004)
\bibitem{scully}
A. Belyanin, V. V. Kocharovsky, F. Capasso, E. Fry, M. S. Zubairy, M. O. Scully, Quantum electrodynamics of accelerated atoms in free space and in cavities, Physical Review A, \textbf{74(2)}, 023807. (2006)
\bibitem{yu-zhang-1}
H. Yu, and J. Zhang, Understanding Hawking radiation in the framework of open quantum systems, Physical Review D, \textbf{77(2)}, 024031. (2008)
\bibitem{huyu-1}
J. Hu, and H. Yu, Entanglement dynamics for uniformly accelerated two-level atoms. Physical Review A, \textbf{91(1)}, 012327. (2015)
\bibitem{yu-zhang-2}
J. Zhang, and H. Yu, Unruh effect and entanglement generation for accelerated atoms near a reflecting boundary. Physical Review D, \textbf{75(10)}, 104014. (2007)
\bibitem{yu-zhang-3}
J. Zhang, and H. Yu, Entanglement generation in atoms immersed in a thermal bath of external quantum scalar fields with a boundary, Physical Review A, \textbf{75(1)}, 012101. (2007)
\bibitem{huyu-2}
J. Hu, and H. Yu, Entanglement generation outside a Schwarzschild black hole and the Hawking effect, Journal of High Energy Physics, \textbf{2011(8)}, 1-13. (2011)
\bibitem{huyu-3}
Y. Yang, J. Hu, and H. Yu, Entanglement dynamics for uniformly accelerated two-level atoms coupled with electromagnetic vacuum fluctuations, Physical Review A, \textbf{94(3)}, 032337. (2016)
\bibitem{oqs-qc-1}
H. Wang, S. Ashhab, and F. Nori, Quantum algorithm for simulating the dynamics of an open quantum system, Physical Review A, \textbf{83(6)}, 062317. (2011)
\bibitem{oqs-qc-2}
A. W. Schlimgen, K. Head-Marsden, L. M. Sager, P. Narang, and D. A. Mazziotti, Quantum Simulation of Open Quantum Systems Using a Unitary Decomposition of Operators, arXiv:2106.12588. (2021)
\bibitem{oqs-qc-3}
J. Han, W. Cai, L. Hu et al, Experimental simulation of open quantum system dynamics via Trotterization, Physical Review Letters, \textbf{127(2)}, 020504. (2021)
\bibitem{oqs-qc-4}
H. Kamakari, S. N. Sun, M. Motta, and A. J. Minnich, Digital quantum simulation of open quantum systems using quantum imaginary time evolution, arXiv:2104.07823. (2021)
\bibitem{martinetti-2}
P. Martinetti, Conformal mapping of Unruh temperature, Modern Physics Letters A, \textbf{24(19)}, 1473-1483. (2009)
\bibitem{casini}
H. Casini, M. Huerta and R. C. Myers, Towards a derivation of holographic entanglement entropy, Journal of High Energy Physics, \textbf{2011(5)}, 036. (2011).
\bibitem{arzano}
M. Arzano, Conformal quantum mechanics of causal diamonds, Journal of High Energy Physics, \textbf{2020}, 1-14. (2020)
\bibitem{birrell-davies}
N. D. Birrell, and P. C. W. Davies, Quantum fields in curved space, Cambridge University Press. (1984)
\bibitem{olson-ralph}
S. J. Olson, and T. C. Ralph, Extraction of timelike entanglement from the quantum vacuum, Physical Review A, \textbf{85(1)}, 012306. (2012)
\bibitem{breuer-book}
H. P. Breuer, and F. Petruccione, The theory of open quantum systems, Oxford University Press. (2002)
\bibitem{nielsen-chuang}
M. A. Nielsen, and I. Chuang, Quantum computation and quantum information, Cambridge University Press. (2002).
\bibitem{kraus-lindblad}
H. Nakazato, Y. Hida, K. Yuasa, B. Militello, A. Napoli, and A.Messina, Solution of the Lindblad equation in the Kraus representation, Physical Review A, \textbf{74(6)}, 062113. (2006)
\end{thebibliography}
\end{document}